\documentclass[useAMS,usenatbib,a4paper]{mn2e}
\usepackage{times}
\title[Giant Molecular clouds]
{
The dependence of stellar age distributions on GMC environment
}
\author[Dobbs]
{C. L. Dobbs\thanks{E-mail:
dobbs@astro.ex.ac.uk}$^{1}$, J. E. Pringle$^2$ and T. Naylor$^1$ \\
$^1$ School of Physics and Astronomy, University of Exeter, Stocker Road, Exeter, EX4 4QL, UK \\
$^2$ Institute of Astronomy, Madingley Road, Cambridge, CB3 0HA }
\usepackage{amssymb}
\usepackage[pdftex]{graphicx}
\usepackage{epsfig}
\usepackage{multirow}

\begin{document}
\label{firstpage}
\date{\today}

\pagerange{\pageref{firstpage}--\pageref{lastpage}} \pubyear{2012}

\maketitle

\begin{abstract}
In this Letter, we analyse the distributions of stellar ages in Giant Molecular Clouds (GMCs) in spiral arms, inter-arm spurs, and at large galactic radii, where the spiral arms are relatively weak. We use the results of numerical simulations of galaxies, which follow the evolution of GMCs and include star particles where star formation events occur. We find that GMCs in spiral arms tend to have predominantly young ($<$ 10 Myr) stars. 
By contrast, clouds which are the remainders of spiral arm GMAs that have been sheared into inter-arm GMCs, contain fewer young ($<$ 10 Myr) stars, and more $\sim20$ Myr stars. We also show that clouds which form in the absence of spiral arms, due to local gravitational and thermal instabilities, contain preferentially young stars. We propose the age distributions of stars in GMCs will be a useful diagnostic to test 
different cloud evolution scenarios, the origin of spiral arms, and the success of numerical models of galactic star formation. We discuss the implications of our results in the context of Galactic and extragalactic molecular clouds.
\end{abstract}


\section{Introduction}
Determining the origin, and formation of observed GMCs is highly challenging. In more recent years, numerical simulations have computed the properties of GMCs in simulations, but whilst \citet{Dobbs2011new} and \citet{Hopkins2011} have cited the role of feedback in determining cloud properties, as yet there are few tests which distinguish the origin of molecular clouds. For nearby clouds, the small range of observed stellar ages, typically several Myr \citep{Jeffries2011,Oliveira2009,Wright2010}, has been used as evidence that these clouds have formed as a result of local converging flows, either due to turbulence, or recent feedback events such as supernovae \citep{Ball1999,Hartmann2012}. However nearby clouds tend to be low mass ($\lesssim 10^4$ M$_{\odot}$), inter-arm clouds, and thus their relevance to the evolution and properties of $10^5-10^6$ M$_{\odot}$ clouds we can resolve in external galaxies, which are though to have longer lifetimes of 20-30 Myr \citep{Kawamura2009} or possibly more \citep{Koda2009}, is unclear. And even when considering clouds of the same mass, some variations are seen across different nearby galaxies \citep{Hughes2013}.

A further complication is the origin of spiral structure in galaxies. If the galaxy has a slowly rotating spiral pattern, e.g. for a quasi-static spiral, or a tidally interacting galaxy, the gas accumulates into GMCs in the arms, and GMCs are sheared out into spurs as they emerge. Such spurs are clearly seen in M51. We also see spurs in the Milky Way, most notably the nearby Orion Spur \citep{Bok1959}. However it is not clear whether this is the same type of feature as seen in M51 or another structure (see. e.g. \citealt{Carraro2013}). If the Milky is a flocculent galaxy, where transient spiral arms are induced by instabilities in the stellar and / or  gas disc, this spur could be a small arm.

In this Letter, we analyse the distribution of stellar ages in clouds from numerical simulations, and relate them to the origin and evolution of the clouds. Here we focus primarily on the simple example of a fixed spiral potential, showing how the ages of stars vary between inter-arm and spiral arm clouds, and the outer parts of the galaxy. In this instance, the most massive clouds are formed by mergers of smaller clouds, and self gravity, and are situated in the spiral arms \citep{Dobbs2008}. The inter-arm clouds, or at least the more massive inter-arm clouds, tend to be spiral arm GMCs that have been sheared out by differential rotation. We also present results from a simulation without spiral arms, where the young stars are situated in clouds formed by local gravitational and thermal instabilities. 

\begin{figure*}
\includegraphics[scale=0.65, bb=30 0 800 410]{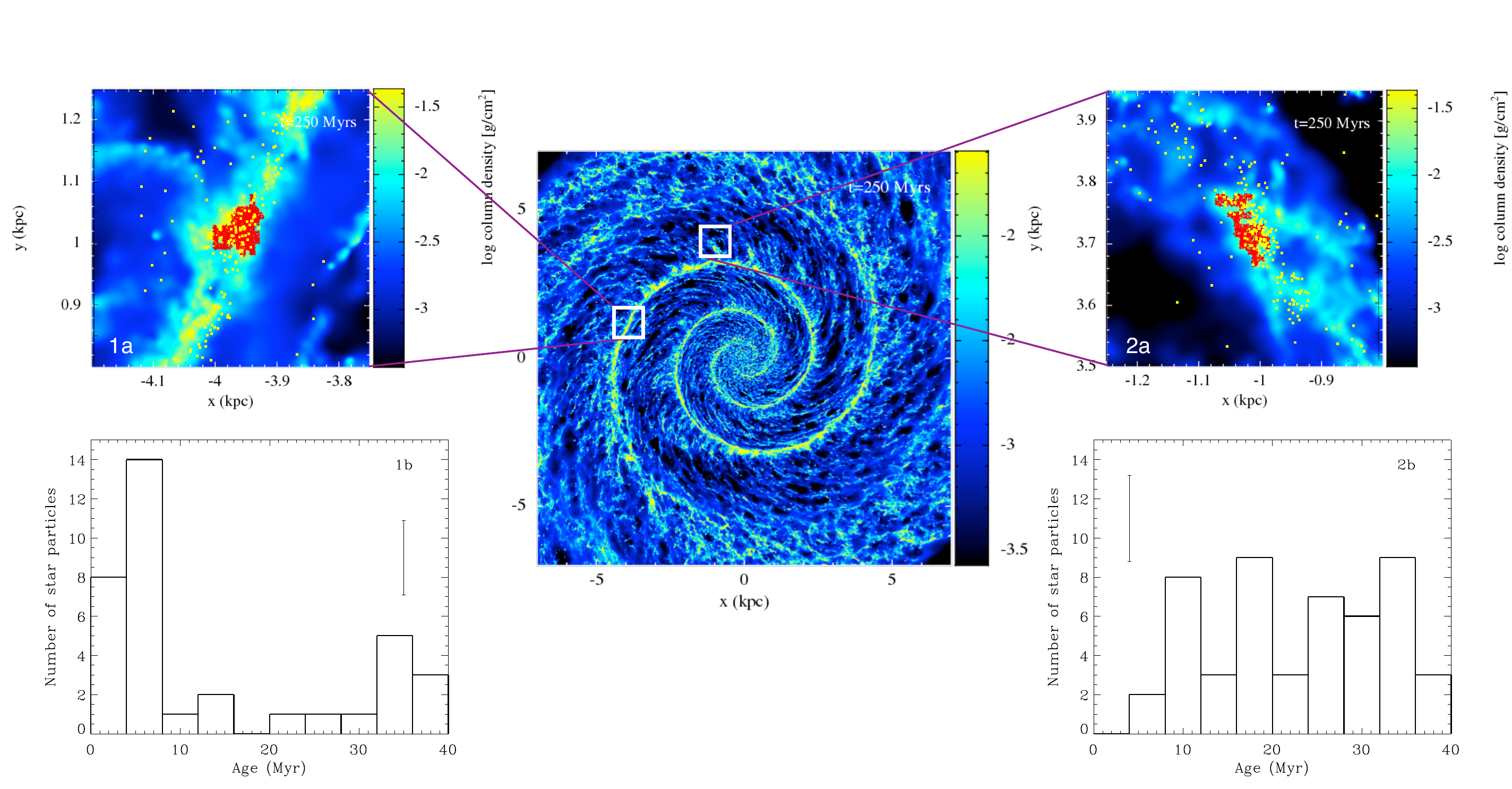}\label{fig:fig1}
\caption{The central panel of Figure~\ref{fig:fig1} shows the galaxy simulation with spiral arms (standard feedback scheme, star particles are not shown on the plot). Panels 1a and 2a show zoomed in regions containing arm and inter-arm clouds, with star particles, each representing 160 M$_{\odot}$ of stars, marked as yellow dots. Panels 1b and 2b show the stellar age distribution for these clouds. Typical error bars, based on the root mean noise are shown on these panels, i.e. $\pm \sqrt{N}$ where N is the average number of particles per bin. The cloud in panels 1a and 1b has a mass of 9 $\times 10^5$ M$_{\odot}$, and the cloud in panels 2a and 2b has a mass of 4.5 $\times 10^5$ M$_{\odot}$.}
\end{figure*}

\section{Numerical calculations}
We perform calculations of galaxies with and without spiral arms. For our calculations with spiral arms, we use a numerical simulation from \citet{Dobbs2013} which modelled a spiral galaxy with a $m=2$ spiral potential, including heating and cooling, stellar feedback and self gravity. The simulation has a gas surface density of 8 M$_{\odot}$ pc$^{-2}$. In this paper we rerun the simulation from a time of 200 Myr, up to around 260 Myr. Most of our analysis is shown at a time of 250 Myr, thus considers stars up to an age of 50 Myr.
We insert feedback (kinetic and thermal energy) each time star formation is assumed to occur (when gas reaches a threshold density) similarly to \citet{Dobbs2013}. In the calculations presented here though, we also convert one particle to a star particle. We calculate the star formation rate by multiplying the mass  of molecular hydrogen involved by an efficiency parameter, here 5\%. In these simulations, like those in \citet{Dobbs2013}, we insert $10^{51}$ ergs (i.e. equivalent to one massive star) for each 160 M$_{\odot}$ of star formation, and we distribute the total energy according to the snowplough phase of a supernova. The mass resolution of our simulation means that each star formation is typically associated with only one feedback event of $10^{51}$ ergs and 160 M$_{\odot}$ of star formation. We then convert one particle (the most dense) to a star tracer particle, which can thus be considered to represent $\sim$160~M$_{\odot}$ of star formation. Thereafter the star particle is subject to the galactic potential, and the gas self gravity.

We ran one simulation where we kept the same feedback prescription as \citet{Dobbs2013} (where feedback was instantaneous), and ran another simulation  where feedback is spread over a time period of 10 Myr (the star particle assigned at the beginning of the 10 Myr period). For the simulation with no spiral arms, we use a smooth galactic potential (thus the resulting structure is similar to Figure~12 of \citealt{Dobbs2011new}). Otherwise the implementation of feedback and star particles is the same. All the simulations contain 8 million particles, and each gas particle has a mass of 312.5~M $_{\odot}$. The simulations initially only contain gas particles, but over the course of the simulation a small fraction ($\lesssim1$ \%) are turned into star particles.

We then apply a clump finding algorithm to locate GMCs in our simulation, as described in previous work \citep{Dobbs2013,Dobbs2008}. Here we adopt a threshold column density of 75 M$_{\odot}$ pc$^{-2}$ to select clouds. This was slightly different to \citet{Dobbs2013}, where we used 100 M$_{\odot}$ pc$^{-2}$ but we chose a slightly lower threshold to ensure getting more massive clouds. For the clouds selected, we can then find the ages of the star particles within the cloud. For all the analysis presented here, we only take clouds which are $>10^5$ M$_{\odot}$  (320 gas particles). We thus use a snapshot of the GMCs in their current state, and the distribution of clusters associated with them, the same information that observers would have about our or another galaxy.

\section{Results}
\subsection{Individual clouds}
In Figure~\ref{fig:fig1} we show results for 2 individual GMCs, using the simulation where we applied the original instantaneous feedback scheme, and included spiral arms. One cloud is situated in a spiral arm, with mass $9\times 10^5$ M$_{\odot}$, and the other in an inter-arm spur, with a mass of $4.5\times 10^5$ M$_{\odot}$. Relatively massive clouds were chosen, but otherwise they are random clouds. The panels labelled b) show the distributions of ages of the star particles. For the spiral arm cloud, there is a clear peak at about 5 Myr, and the cloud contains mostly young stars. For the cloud which lies in a spur, there are no young stars of $<$ 5 Myr. Instead, the most common age of stars is $\sim$ 18 Myr, but with a broad distribution from 5-40 Myr. The inter-arm cloud also has relatively more stars per unit mass, reflecting that the cloud is at a later stage of its lifetime and exhibiting a higher star formation efficiency. That this particular cloud has no $<$ 5 Myr star particles seems largely random - as we show in the next section, some inter-arm clouds do have young stars, although there is generally a scarcity of such stars compared to older populations.

For comparison, clouds stay in the arms in these simulations for about 30-40 Myr before they are sheared out (see \citealt{Dobbs2013}, Figure~3). Thus the gas in inter-arm clouds is likely to have entered a spiral arm 40-50 Myr earlier (on average, as some gas spends a relatively short amount of time in the arm, and some longer), and indeed by tracing the gas back in time, we find this is the case for the inter-arm cloud shown in Figure~\ref{fig:fig1}.

\subsection{Results summed over all clouds in different regions}
\begin{figure*}
\centerline{\includegraphics[scale=0.35]{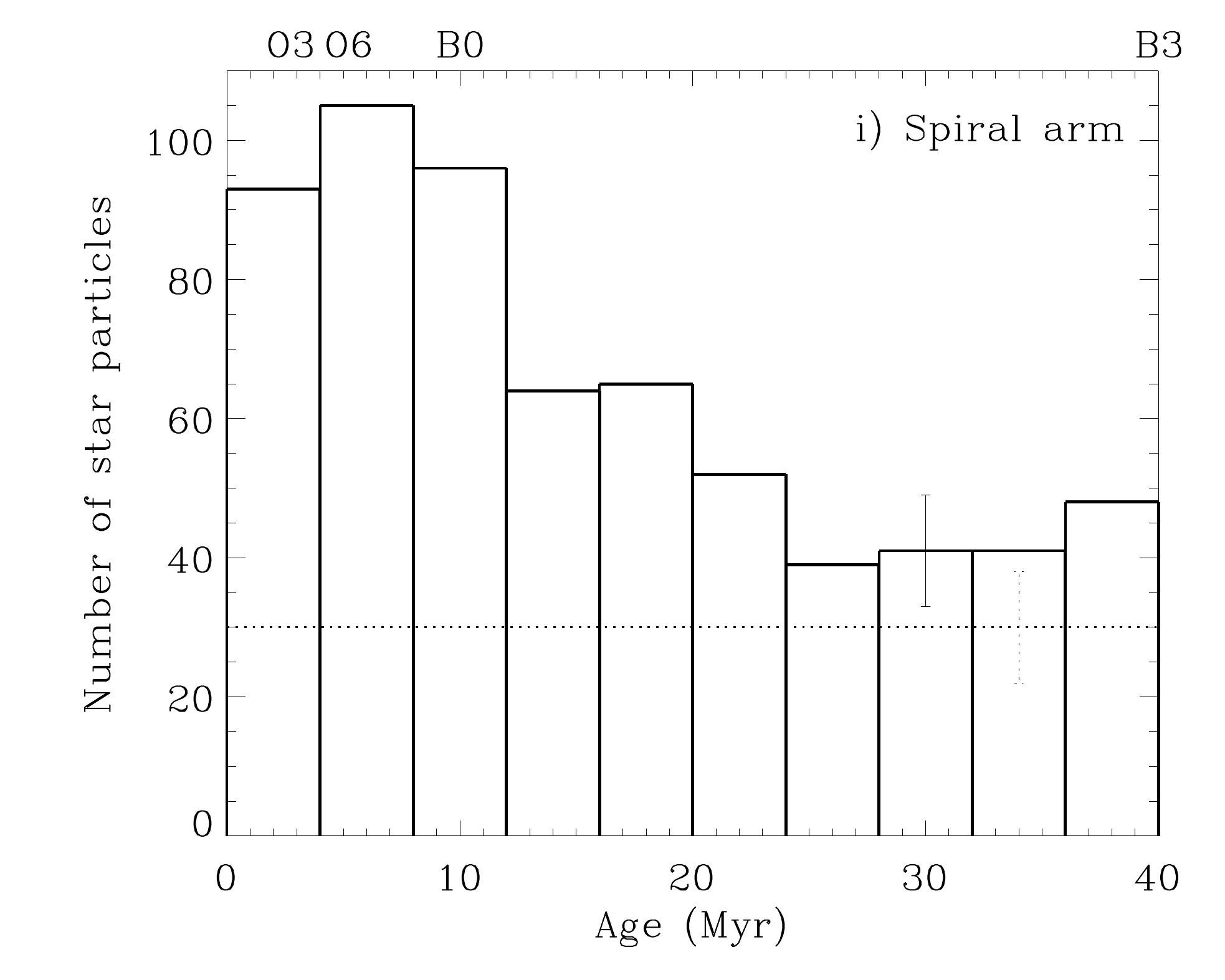}
\includegraphics[scale=0.35]{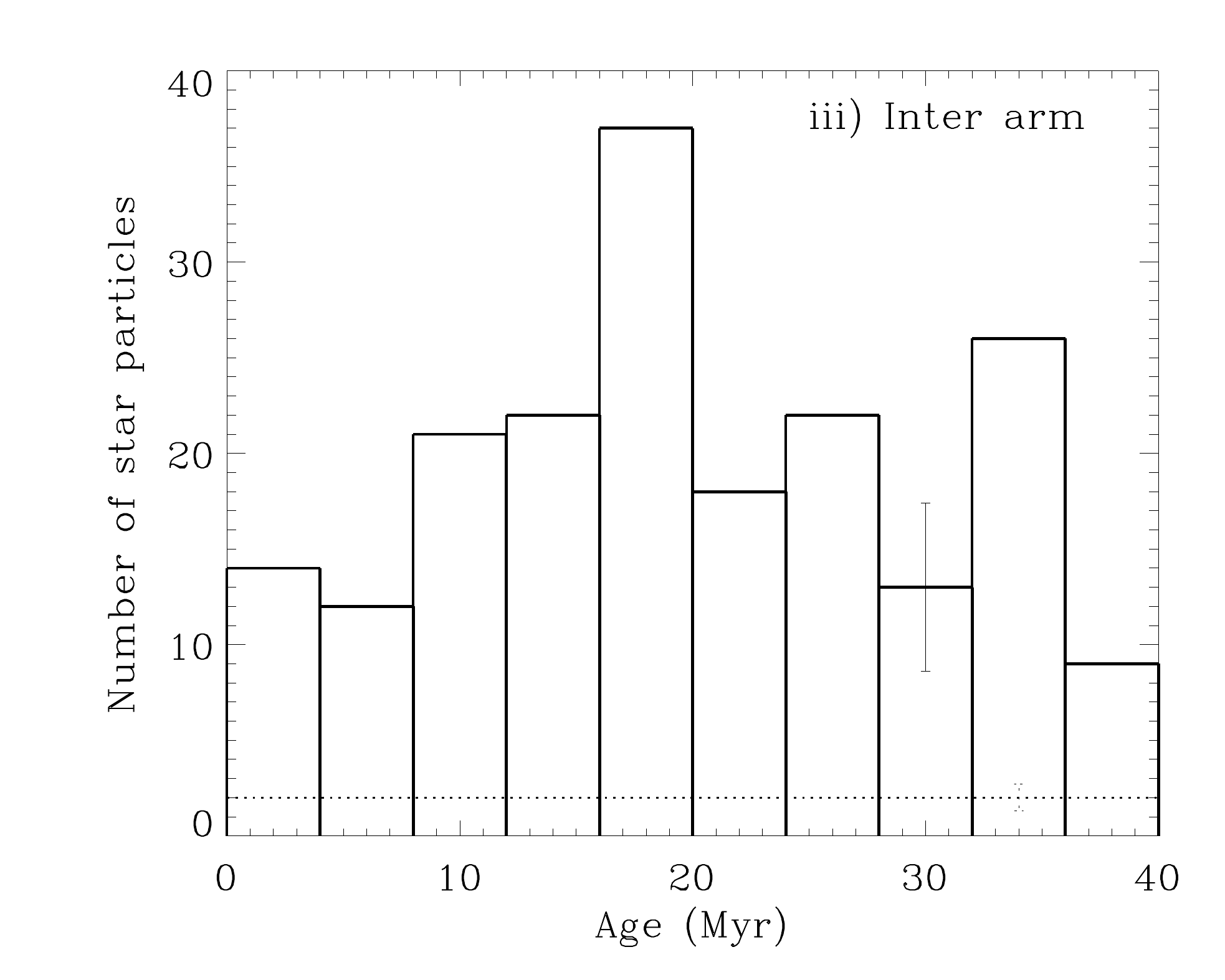}}
\centerline{\includegraphics[scale=0.35]{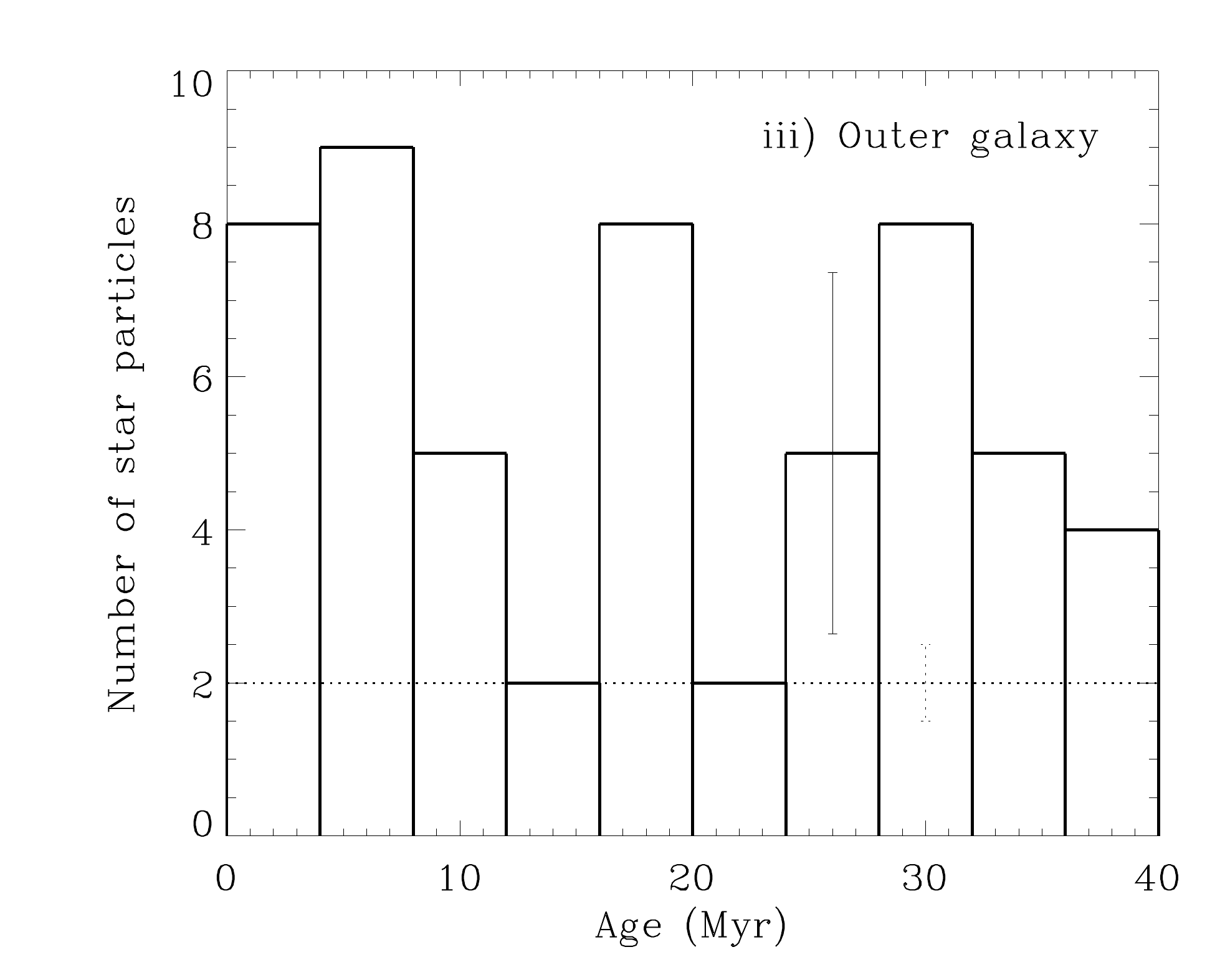}
\includegraphics[scale=0.35]{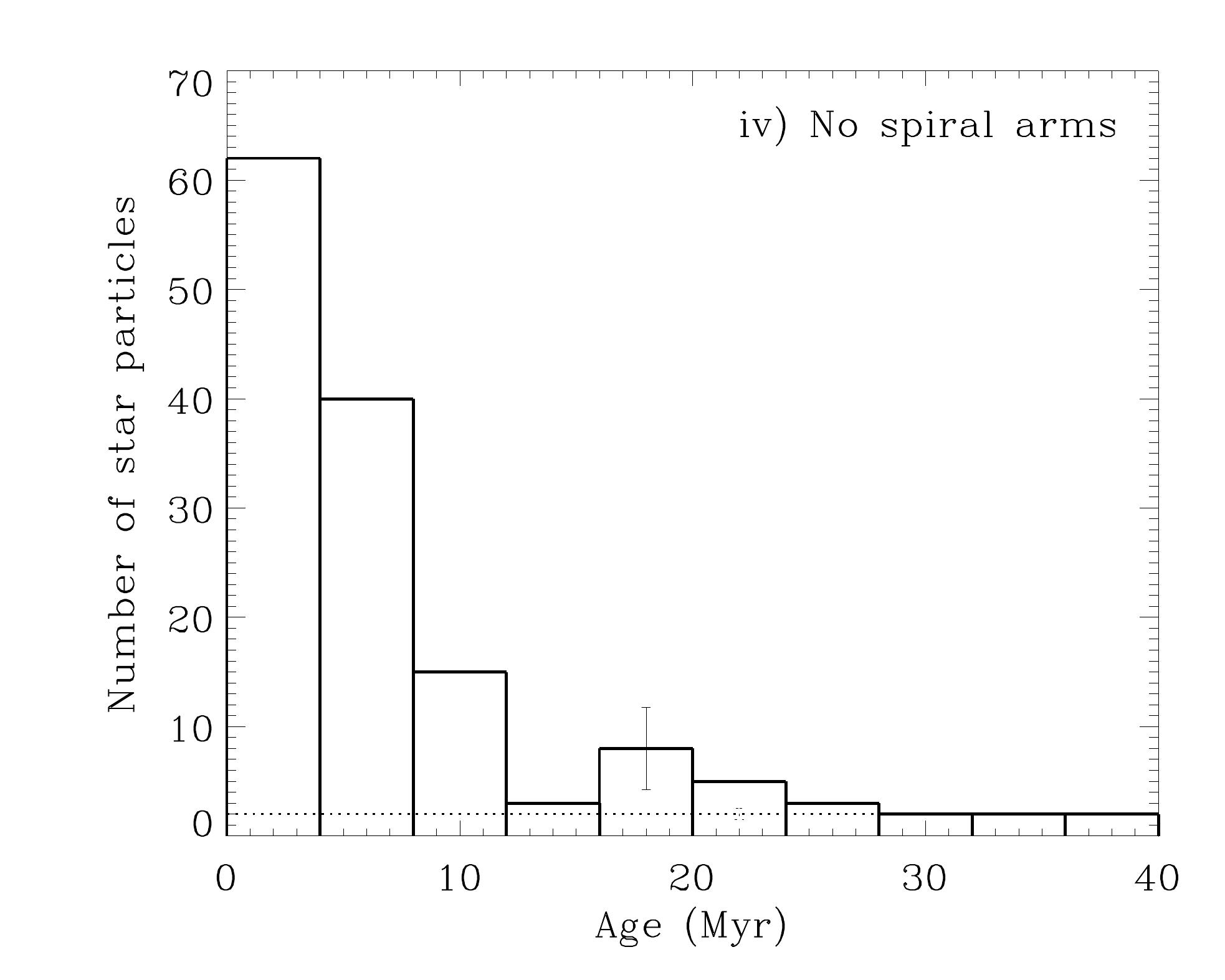}} 
\caption{Stellar age distributions are plotted for clouds in different environments. Panels i-iii) use a simulation with an imposed spiral potential, whereas in iv) the simulation has no imposed spiral potential. In each case, the stellar age distributions are summed over all clouds over $10^5$ M$_{\odot}$. Typical error bars, based on the root mean noise are shown on each panel. The dotted lines show the expected background level for each environment, assuming a constant star formation rate (which is reasonable for the 40 Myr period we consider here), with associated error bars. The error bars for the background level incorporate uncertainties in measuring the areas of clouds, and estimating differences between the spiral arms arm and inter-arm regions. Also shown across the top of panel i) are approximate lifetimes of different spectral types \citep{Massey1995,Buzzoni2002,Martins2005,Hohle2010}. There are clearly differences in the stellar age distributions in different environments. The clouds in the spiral arms, and those which form in the simulation without a spiral potential (where clouds will primarily form by gravitational and thermal instabilities) preferentially contain young stars. The inter-arm clouds, however contain a broad distribution, about an age of $\sim$ 20 Myr.}
\label{fig:fig2}
\end{figure*}

To improve our statistics, we compute the age distributions for all $>10^5$ M$_{\odot}$ clouds in our original spiral galaxy calculation (shown in Figure~\ref{fig:fig1}). We divided our sample into 3 regions, spiral arms with $r<6$ kpc, inter-arm clouds with $r<6$ kpc, and outer arm clouds, with $r>6$ kpc (containing 68, 12 and 26 clouds respectively). A distinction between inner and outer regions was made at 6 kpc, because as seen in Figure~\ref{fig:fig1}, the spiral structure is much stronger in the inner regions. The inter-arm clouds were selected by eye from the inner region clouds. In Figure~\ref{fig:fig2} we show the stellar age distributions summed over all the ($>10^5$ M$_{\odot}$) clouds in the arm and inter-arm regions.  In Figure~\ref{fig:fig2} we also plot a background level of stars, the estimated average number of stars in each bin for each environment,
assuming a time independent star formation rate. We determine the background level of star formation by multiplying the total number of stars formed by an estimate of the fraction of gas in that environment (detailed processes such as shear are neglected, but we do allow for regions with higher or lower surface densities to exhibit linearly higher or lower star formation rates).
Figure~\ref{fig:fig2} shows similar
distributions to those for the individual clouds shown in Figure~\ref{fig:fig1}. A KS test confirms that, with a p-value of $1.7\times10^{-6}$, the distributions for the arm and inter-arm clouds are statistically different. Figure~\ref{fig:fig2} also shows the distribution for the outer region of the galaxy, which is somewhat more random. This could be because there is still an influence of the spiral potential at these larger radii, and that some clouds have been in the minimum of the potential for longer than others and have different distributions of stars. There were also relatively fewer star particles at the larger radii making this analysis more difficult. KS tests gave p-values of 0.1 and 0.15 compared with the arm and inter-arm GMCs respectively. Thus the outer arm clouds could not be statistically distinguished from the arm or inter-arm clouds.  

We also tested whether the distributions in Figure~\ref{fig:fig2} were statistically different from a uniform distribution. The distribution for the arm clouds is statistically different (p$=$0.03), whereas the outer galaxy and inter-arm clouds are not (p$=$0.3 and 0.1 respectively). The inter-arm clouds are likely formed by a combination of smaller clouds of different ages, as well as some small amount of ongoing star formation, so exhibit a more uniform distribution of ages, compared to the spiral arm clouds.  

We also examined the stellar age distributions for clouds according to their mass. We divided the clouds into those with fewer than 1000 particles ($<3.12\times 10^5$ M$_{\odot}$), and those with more than 1000 particles ($>3.12\times 10^5$ M$_{\odot}$). For the spiral arms clouds, the distribution for the massive clouds was less noisy, and had a peak at 4-6 Myr, whereas the peak for the low mass clouds was 6-10 Myr. Overall though the distributions were not significantly different.
For the inter-arm clouds, the statistics become too small to say anything meaningful, although there was no obvious distinction between the two ranges in mass. 

\subsection{Results for different times, and alternative feedback scheme}
We also computed the stellar age distributions in clouds at times of 240 and 260 Myr, and repeated our analysis for a simulation where we spread feedback over a period of 10 Myr. However in all cases, we found similar age distributions to those shown in Figure~\ref{fig:fig2}. Some cases were more noisy, some less, but adding all distributions together still shows a  strong bias towards  0-10 Myr age stars for the arm clouds, and $\sim20$ Myr age stars for the inter-arm clouds. When taking a larger sample of outer galaxy clouds (by summing over multiple time frames), we see a tendency for a clearer decline in the stellar age distribution compared to Figure~\ref{fig:fig2} panel iii). However there is still not a distinct difference to either the arm, or inter-arm clouds, the ages seeming more an average of the two distributions.

\subsection{Results for galaxy with no imposed spiral potential}
Finally, in Figure~\ref{fig:fig2} panel iv), we show the stellar age distribution summed over all the $>10^5$ M$_{\odot}$ clouds (32 in total) in the simulation without a spiral potential. In this case, the stellar ages are clearly predominantly $\lesssim 10$ Myr, and though there are a few older stars present, their number is no higher than the level of background stars expected. We also tested whether the width of the distribution was statistically different to that for the distribution of GMCs in spiral arms. We fitted a gaussian distribution (with a constant background level of star particles) to each of the distributions in Figure~\ref{fig:fig2} panels i) and iv), and carried out an f test to check whether the variances were statistically different. They are statistically different, the distribution of stellar ages in the no spiral arm case being significantly narrower, when fitting half gaussians and adopting  the peaks in panels i) and iv) as the means (for full gaussians, the variances are not statistically different, but the fit for panel iv) gives an unrealistic negative mean). 

\subsection{Interpretation}
For the spiral galaxy, the GMCs in the spiral arms have formed relatively recently, and thus contain young stars. The inter-arm clouds contain stars that were formed when that gas was last in the spiral arms. The age gradient in stars going away from the arm is similar to the predictions of stellar ages shown in \citet{Dobbs2010b}. In the spiral arms, clouds which are merging due to converging orbits in the spiral shock, or being formed by gravitational instabilities, are the sites of converging flows \citep{Dobbs2012}, hence are actively forming stars. This is in contrast to clouds in the inter-arm regions, which are being sheared out rather than undergoing compression. Hence these inter-arm clouds contain relatively fewer (though still some) younger stars. 

For the symmetric galaxy, without stellar spiral arms, the GMCs form by gravitational and thermal instabilities only. The clouds do not reach particularly high masses and are readily dispersed by feedback. Hence these clouds tend only to be young, and contain young stars.

\subsection{Discussion}
Our analysis of the stellar age distributions suggests a way of distinguishing between clouds which have formed \textit{in situ}, i.e. in their present environment, or instead are remnants of spiral arm clouds which are now in the inter arm regions. Clouds which have formed in their present environment could be spiral arm clouds formed by agglomeration and self gravity, or clouds formed by gravitational instabilities or otherwise in the absence of spiral arms. Our results are probably most applicable for comparison with extragalactic clouds, but we consider them in the context of Galactic clouds too, where age distributions are observed. 

Observations of GMCs in other galaxies indicate that they have lifetimes of 20--30 Myr, which if true, would be sufficiently long enough to test whether there are differences in the age distributions depending on GMC environment. In particular, galaxies which have clear inter-arm spurs, would provide a good test for comparisons with these simulations. We note that our models do not say there are no spiral arm clouds with an older stellar population, or inter-arm clouds with many young stars, but statistically, the presence of such clouds is much less likely. 

Searching for observational evidence for a difference in age spread between star-forming regions is complicated by the fact that at a given effective 
temperature the pre-main-sequences of all young clusters show a spread in luminosity. Whilst the simplest explanation of this is an age spread, many other explanations are possible \citep{Soderblom2013}.  This means we must discount the age spreads found within individual clusters.
However these observations typically correspond to regions of mass at most a few $10^4$ M$_{\odot}$, whereas for this paper we are interested in stellar ages over entire $10^5-10^6$ M$_{\odot}$ complexes, containing multiple clusters. Some such measurements are available for a few more massive clouds. Clusters associated with the Orion complex, which has a mass of $\sim 2\times 10^5$ M$_{\odot}$ and is situated along the Orion spur, exhibit ages from $\lesssim$2 Myr up to $\sim12$ Myr \citep{Briceno2008,Bally2008,DaRio2010,Reggiani2011}. 
The nearby Sco-Cen association, which is also in an inter-arm region, shows an age spread of $\sim15$ Myr \citep{Lawson2001,Mamajek2002,Feigelson2003}.
Clusters in the Carina nebula, which is $\sim 7\times 10^5$ M$_{\odot}$ and located in a spiral arm, exhibit ages up to $\sim 10$ Myr \citep{Smith2008,Townsley2011}. Although statistically we do not have a large sample of Milky Way objects, these age spreads are consistent with our simulations, and both the simulations and observations indicate a genuine age spread in the stars. However the observed age spreads are least consistent with those we find for our simulated inter-arm spur clouds, suggesting perhaps the Galactic clouds all formed \textit{in situ} (and were not previously spiral arm clouds). Such a scenario could perhaps be more consistent with the Milky Way being a more flocculent spiral galaxy, although we emphasise that we have only considered the age distributions of three, nearby GMCs. The shapes of the observed age distributions also tend to be roughly Guassian, which is most consistent with the spiral arm clouds, and the clouds in the simulation with no spiral arms.

In external galaxies there is the potential to measure age spreads for large star-forming complexes using upper main-sequence and post-main-sequence stars. Within the local group such stars can be resolved by HST observations, allowing them to be placed on a Hertzsprung-Russell or colour-magnitude diagram, and their ages derived from isochrones (e.g \citealt{Bianchi2012}). Such ages are much less prone to error than pre-main-sequence ages \citep{Naylor2009} and so detecting a spread in age over a star-forming complex could be a sensitive diagnostic. Integrated-light observations would be more problematic, since the model fitting is already often degenerate in two or more parameters, a situation epitomised by the range of possible explanations for discrepancies between H$\alpha$ data and derived bolometric luminosities (e.g. \citealt{Grossi2010}).

In this paper, we have considered two extreme cases: a model with a fixed spiral pattern, and a model with no stellar spiral arms, where the structure is only present in the gas. For the latter case, no clouds with an older stellar population were found. We did not consider the case of galaxies with transient stellar arms. However previous work has indicated that in these galaxies, clouds disperse as the spiral arm disperses, and inter-arm spurs do not form in the same way \citep{Dobbs2007,Wada2011}. Thus inter-arm clouds are likely to form by gravitational instabilities, similarly to are model with no stellar arms, and so in these, transient spirals, we would also expect clusters to contain predominantly young stars.

Finally we have not included cluster dispersion in our models, rather we simply have particles that represent clusters. N-body effects, as well feedback, may  disperse the cluster away from the molecular clouds. We are also unable to tell at which stage our star particles would disperse into field stars.  Much higher resolution, or zoom in simulations, would be needed to study cluster evolution in conjunction with the evolution of the clouds.
Nevertheless we present here a first analysis of stellar ages in GMCs in a galactic context.

\section{Acknowledgments}
We thank the referee for a constructive and thoughtful report.
The calculations for this paper were performed primarily on the DiRAC machine `Complexity', as well as the supercomputer at Exeter, which is jointly funded by STFC, the Large Facilities Capital Fund of BIS, and the University of Exeter. 
CLD acknowledges funding from the European Research Council for the 
FP7 ERC starting grant project LOCALSTAR. Figure~\ref{fig:fig1} was produced using \textsc{splash} \citep{splash2007}.
\bibliographystyle{mn2e}
\bibliography{Dobbs}

\bsp
\label{lastpage}
\end{document}